\documentclass[10pt, a4paper]{article}
\usepackage[final]{lrec-coling2024} 
\usepackage[framemethod=TikZ]{mdframed}
\usepackage{booktabs}

\usepackage{booktabs}
\definecolor{best}{rgb}{0.267, 0.447, 0.769}

\usepackage{amsthm,amsmath,amssymb,amsfonts}

\newtheorem{proposition}{Proposition}
\newtheorem{definition}{Definition}[section]

\title{Can ChatGPT Compute Trustworthy Sentiment Scores from Bloomberg Market Wraps?} 
\name{B. Lefort\textsuperscript{1,2}, E. Benhamou\textsuperscript{1,3}, JJ. Ohana\textsuperscript{1}, D. Saltiel\textsuperscript{1}, B. Guez\textsuperscript{1}, D. Challet\textsuperscript{2}}

\address{\textsuperscript{1}: Ai for Alpha, \textsuperscript{2}: CentraleSupelec, \textsuperscript{3}: Dauphine PSL \\
         \{baptiste.lefort, eric.benhamou, jean-jacques.ohana, david.saltiel, beatrice.guez\}@aiforalpha.com \\
         damien.challet@centralesupelec.fr\\}

\abstract{
We used a dataset of daily Bloomberg Financial Market Summaries from 2010 to 2023, reposted on large financial media, to determine how global news headlines may affect stock market movements using ChatGPT and a two-stage prompt approach. We document a statistically significant positive correlation between the sentiment score and future equity market returns over short to medium term, which reverts to a negative correlation over longer horizons. Validation of this correlation pattern across multiple equity markets indicates its robustness across equity regions and resilience to non-linearity, evidenced by comparison of Pearson and Spearman correlations. Finally, we provide an estimate of the optimal horizon that strikes a balance between reactivity to new information and correlation.\\
\newline 
}

\begin{document}
\maketitleabstract

\section{Introduction}\label{sec:Introduction}
Finance has a longstanding tradition of employing Natural Language Processing (NLP) to extract valuable insights from textual data and news \citep{Tetlock2007GivingMarket,schumaker2009textual}. The financial world has always been at the forefront of embracing technological innovation. From the inception of electronic trading to the burgeoning realm of fintech, financial services have undergone significant evolution, especially with the arrival of AI and ML technologies \citep{arner2015evolution, fatouros2023deepvar}.

Sentiment analysis stands out as a cornerstone in this transformation \citep{poria2016aspect}. It plays a crucial role in deciphering market sentiments, offering invaluable predictive insights. Historically, the financial sector leaned on handpicked word lists and basic ML techniques for sentiment analysis \citep{Tetlock2007GivingMarket,schumaker2009textual}. Yet, with NLP's rapid advancements, a slew of advanced methods has come to the fore. Models like BERT and its finance-centric sibling, FinBERT, have elevated sentiment analysis's precision \citep{devlin2018bert, liu2021finbert}.

However, the financial realm brings its set of challenges for sentiment analysis \citep{loughran2011liability}. Financial news is a complex mesh of domain-specific jargon and layered emotions. A singular piece of news might carry different sentiments for multiple financial entities, making general sentiment analysis tools potentially misleading. News may also come after the facts and hence have no real predictive power. Furthermore, these tools often struggle with context-specific outputs, making them less versatile in diverse scenarios \citep{poria2017review}. Indeed, undertaking natural language processing (NLP) in finance is notably challenging due to the specificity of the corpus, as evidenced by diverse studies on financial texts, sentiment lexicons, and financial reports across various languages and financial systems \citep{li2022finmath, moreno2020design, ghaddar2020sedar} and can require knowledge graph \citep{oksanen2022graph} or language-specific corpus \citep{masson2020nlp, jabbari2020french, zmandar2022cofif}. Converting a sentiment score into an investment strategy is notably difficult \citep{yuan2020target,iordache2022investigating}

With the advent of Large Language Models (LLMs), an AI paradigm has emerged with transformative potential \citep{george2023review}. GPT, particularly its conversational variant, ChatGPT, has shown promise in refining financial applications \citep{openai2023gpt4}. By leveraging ChatGPT's prowess in language comprehension, financial entities can enhance their sentiment analysis depth. This proficiency translates to better-informed investment decisions, optimized risk management, and more effective portfolio strategies. Furthermore, ChatGPT's capability to convey intricate financial insights in understandable terms makes it a potential game-changer in democratizing financial knowledge \citep{yue2023democratizing}.

In this study, we design a sentiment analysis of Bloomberg markets wrap news using ChatGPT. Besides, we developed a two-step prompt-based process to extract information from text and convert this into a sentiment score. Finally, we show that this score enables us to understand better the effect of the news on the market especially regarding cyclic and counter-cyclic behavior. To sum up, the contributions of this paper are three folds:
\begin{enumerate}
    \item We designed a two-step ChatGPT based sentiment analysis extraction from  Bloomberg markets wrap news.
    \item We proposed an index for assessing the ability of ChatGPT to give a sentiment to the news.
    \item We demonstrated that this score reveals significant insights into market behavior and possesses robust predictive capabilities.
\end{enumerate}

The rest of this paper is organized as follows. Section \ref{sec:Related works} briefly reviews the related works. 
Section \ref{sec:Prompt engineering} describes our prompt design and explains how using a two-step method for creating prompts can lead to better sentiment scores than using a one-step approach. Section \ref{sec:Global Equities Sentiment Indicator} outlines the methodology for calculating the sentiment score. Section \ref{sec:Evaluation of the Sentiment Score's Validity} evaluates the sentiment score validity. Section \ref{sec:Trade-Off Analysis of Financial Indicators} discuss the trade-off between using short term predictions with lower correlation or longer period prediction but with the disadvantage of slow reaction to new informations. Section \ref{sec:Robustness over the Equities Markets} review its robustness across various markets. Finally Section \ref{sec:Conclusion} concludes.

\section{Related works}\label{sec:Related works}
In the realm of finance and economics, several recent scholarly works have employed ChatGPT, such as \citet{Hansen2023CanFedspeak}, \citet{Cowen2023HowGPT}, \citet{Korinek2023LanguageResearch, Lopez-Lira2023CanModels}, and \citet{Noy2023ExperimentalIntelligence}. \citet{Hansen2023CanFedspeak} elucidates how Large Language Models (LLMs) like ChatGPT can decipher Fedspeak, the nuanced language employed by the Federal Reserve to convey monetary policy decisions. \citet{Lopez-Lira2023CanModels} explains proper prompting for forecasting stock returns. Both \citet{Cowen2023HowGPT} and \citet{Korinek2023LanguageResearch} elaborate on ChatGPT's utility in economics education and research. Meanwhile, \citet{Noy2023ExperimentalIntelligence} underscores ChatGPT's capability to augment productivity in professional writing tasks. Furthermore, \citet{Yang2023LargeCredibility} showcases ChatGPT's aptitude for distinguishing credible news outlets. 

Simultaneously, research by \citet{Xie2023TheChallenges} posits that ChatGPT's performance is comparable to rudimentary methods like linear regression for numerical data-based prediction tasks. Additionally, \citet{Ko2023CanPerspective} endeavoured to employ ChatGPT in portfolio selection, albeit without discernible success. Our hypothesis attributes these varied outcomes to their reliance on historical numerical data for prediction, whereas ChatGPT's forte lies in textual tasks.

Our paper offers a novel perspective on this body of literature. It pioneers the assessment of ChatGPT's proficiency in forecasting the trends in the NASDAQ, a pivotal task for which it has not been explicitly trained, traditionally referred to as zero-shot learning. Instead of leveraging finance-specific data, we hinge on ChatGPT's intrinsic NLP capabilities. Moreover, we introduce an innovative prompting method to leverage ChatGPT's analytical processes by finding headlines, then converting these headlines into a sentiment, and finally aggregating carefully these scores with both a cumulated sum and a detrended process to filter out noise. Such insights not only augment the nascent literature on deciphering intricate news with LLM models but also differentiate our study from contemporaneous works that use chatGPT in a more brute-force way.

\section{Prompt engineering}\label{sec:Prompt engineering}
\subsection{Data collection}
We collected Bloomberg Global Markets Wrap summaries from 2010 to October 2023. We ignored any text that is less than 600 characters long or any news summary that is not explicitly a market wrap by removing any text that does not contain the keywords "market(s) wrap". Over 3600 news items were collected for applying a two-step approach detailed in section \ref{twostepapproach}. Considering that these summaries encapsulate daily market developments across 10 to 20 headlines, the aggregate dataset is indicative of 36 to 72 thousand comprehensive news items, meticulously curated and verified.

\subsection{Two-step approach}\label{twostepapproach}
We opted to decompose the instructions into simpler and more straightforward tasks. In accordance with the recommendations posited in \cite{Lopez-Lira2023CanModels}, we devised two prompts to refine the objectives for ChatGPT, focusing on tasks empirically demonstrated to align well with ChatGPT's capabilities. Our first prompt consisted of summarizing the text into titles or headlines as follows:\\

\noindent \textbf{First Prompt:}

\begin{mdframed}[linecolor=white, innerleftmargin=10pt, innerrightmargin=10pt]
\textit{Assume you are an experienced asset manager. Analyze the text between \{\} and identify the predominant themes. For each theme, formulate a compelling headline that encapsulates its core message. Please arrange your responses in a list format, ensuring a line break after each headline. \\ Your list should contain a total of 15 distinct headlines reflecting the respective themes and presented in the following format:\\
 1. Headline that encapsulates Theme 1\\
 2. Headline that encapsulates Theme 2\\
...\\
15. Headline that encapsulates Theme 15\\
\{INSERT\_TEXT\_HERE\}}
\end{mdframed}
\vspace{0.2cm}

Our second prompt consisted of determining a sentiment score on each headline: \\
\noindent \textbf{Second Prompt:}
\begin{mdframed}[linecolor=white, innerleftmargin=10pt, innerrightmargin=10pt]
\textit{Assume you are an experienced asset manager. Your task is to assess the impact of various economic events and trends on global equities. For each numbered statement provided below between\{\}, classify its impact as either "positive," "negative," or "indecisive".
\{INSERT\_TEXT\_HERE\}}
\end{mdframed}

For the two prompts, we used the gpt-4.0 version of ChatGPT. The overall idea of this two-step approach is to ease the task of chatGPT and leverage its capacity to make summaries and in a second step find the tone or sentiment. We can now devise an enhanced and more pertinent "Global Equities Sentiment Indicator".

\section{Global Equities Sentiment Indicator}\label{sec:Global Equities Sentiment Indicator}

\begin{definition}\textbf{Daily Sentiment Score:}\label{def:dailysentimentscore}
Let us denote \( h_i \) as the \( i^{th} \) headline scanned from the daily news $n$ and have two scoring functions that are consistent, a positive one \( p(h_i) \) which returns 1 if \( h_i \) is positive, 0 otherwise and a negative one \( n(h_i) \) which returns 1 if \( h_i \) is negative, 0 otherwise.

The sentiment score \( S \) for a day with \( N \) headlines is given by:
\begin{equation}
S = \frac{\sum_{i=1}^{N} p(h_i) - \sum_{i=1}^{N} n(h_i)}{\sum_{i=1}^{N} p(h_i) + \sum_{i=1}^{N} n(h_i)} 
\end{equation}
\end{definition}

The sentiment score \( S \) measures the relative dominance of positive versus negative sentiments in a day's headlines. It satisfies a couple of simple properties that are trivial to prove. As described in table \ref{tab:sentiment}, once we have the daily individual positive and negative score, the sentiment score is easily computed. Moreover, the sentiment score satisfies some properties as highlighted in proposition  \ref{prop:canonical properties}.

\begin{proposition}\label{prop:canonical properties}
The sentiment score \( S \)  satisfies some properties:
\begin{enumerate}
    \item \textbf{Boundedness}: \( S \) is bounded as \( -1 \leq S \leq 1 \).
    \item \textbf{Symmetry}: If sentiments of all headlines are reversed, then \( S \) changes its sign.
    \item \textbf{Neutrality}: \( S = 0 \) if there are equal numbers of positive and negative headlines.
    \item \textbf{Monotonicity}: \( S \) increases as the difference between positive and negative headlines increases.
    \item \textbf{Scale Invariance}: \( S \) remains the same if we multiply the number of both positive and negative headlines by a constant.
    \item \textbf{Additivity}: The combined \( S \) for two sets of headlines is the weighted average of their individual \( S \) values.
\end{enumerate}
\end{proposition}

\begin{table}[!htbp]
    \centering
    \resizebox{\columnwidth}{!}{%
    \begin{tabular}{|c|c|c|c|}
        \hline
        \textbf{Date} & \textbf{Positive} & \textbf{Negative} & \textbf{Score} \\
        \hline
        2010-01-04 & 11 & 3 & 0.57\\
        \hline
        2010-01-05 & 6 & 6 & 0.00 \\
        \hline
        \multicolumn{4}{|c|}{\centering\ldots} \\
        \hline
        2023-11-21 & 8 & 3 & 0.45 \\
        \hline
    \end{tabular}
    }
    \caption{Sentiment Analysis Dataset}
    \label{tab:sentiment}
\end{table}

\begin{figure}[!htbp]
\begin{center}
\includegraphics[width=\columnwidth, height=2cm]{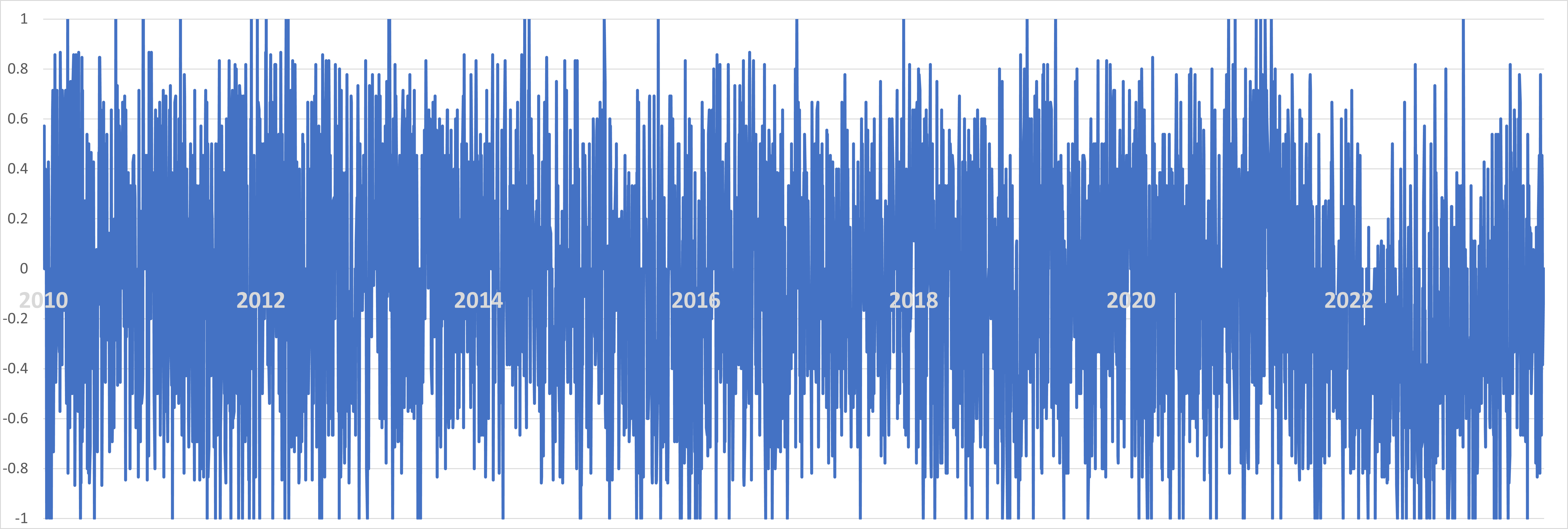}
\caption{Raw signal exhibiting significant noise}
\label{fig:raw_signal}
\end{center}
\end{figure}

Figure \ref{fig:raw_signal} depicts the raw signal corresponding to the score, which exhibits significant noise. Using raw sentiment scores from daily news headlines often results in noisy and less interpretable outcomes. To address this, we propose a \textit{cumulated sentiment score} over a specified period. This score aggregates news sentiments over a duration, offering a more comprehensive measure of the news impact during that period. 

\begin{definition}\label{def:cumulative_sentiment_score}
\textbf{Cumulated Sentiment Score:}We defined a cumulative score as follows. Given:
\begin{itemize}
    \item \( h_{i,t} \) as the \( i^{th} \) headline on day \( t \).
    \item \( p(h_{i,t}) \) and \( n(h_{i,t}) \) as functions returning 1 for positive and negative sentiments of \( h_{i,t} \) respectively, 0 otherwise.
    \item \( d \) as the duration.
\end{itemize}
The cumulated sentiment score \( S_d \) over period \( d \) is:
\begin{equation}
S_d = \frac{\sum_{t=1}^{d}\sum_{i=1}^{N_t} p(h_{i,t}) - \sum_{t=1}^{d}\sum_{i=1}^{N_t} n(h_{i,t})}{\sum_{t=1}^{d}\sum_{i=1}^{N_t} p(h_{i,t}) + \sum_{t=1}^{d}\sum_{i=1}^{N_t} n(h_{i,t})} 
\end{equation}
with \( N_t \) being the number of headlines on day \( t \).
\end{definition}

\begin{figure}[!htbp]
\begin{center}
\includegraphics[width=\columnwidth, height=2.5cm]{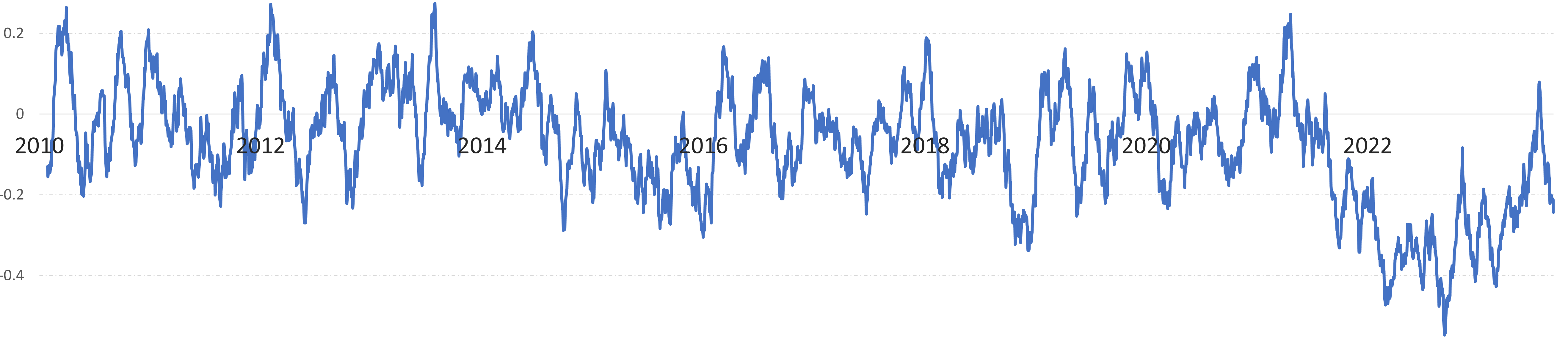}
\caption{Cumulated sentiment score with d=20}
\label{fig:monthly_signals}
\end{center}
\end{figure}

The mathematical properties of proposition \ref{prop:canonical properties}, that is boundedness, symmetry, neutrality, monotonicity, scale invariance remains for the cumulated sentiment score. Figure \ref{fig:monthly_signals} illustrates how the cumulated process diminishes the noise within the signal. \\

The cumulative sentiment enabled us to obtain the trend of the news rather than a momentary snapshot of it, which appeared to be informative.

\section{Evaluation of the Sentiment Score's Validity}\label{sec:Evaluation of the Sentiment Score's Validity}
\subsection{Descriptive statistics}
In order to evaluate the performance of our sentiment score to reveal information about the market reaction, we consider two correlation metrics: Pearson and Spearman coefficient as presented in \cite{wilcox2010fundamentals}. While Pearson correlation coefficients capture linear relationship, the Spearman rank correlation coefficients are a measure of the  monotonic relation between the two variables thanks to the ordering of the rank functions and can deal with ordinal or non-normally distributed data, providing a robust measure of association for non linear data.

\subsection{The Equity Data and Variable Computation}
To assess the robustness of the score, we computed its correlation with diverse equity markets: the SP 500, NASDAQ 100, Nikkei 225, Eurostoxx 50, FTSE 100, and MSCI Emerging Countries indices. We call these markets respectively US, US Tech, Japan, Europe, UK and Emerging equities markets or simply by their region without mentioning equities market explicitly. We used data from January 2010 to November 2023 and computed the resulting returns over multiple periods \( (p_i)_{i=1..n} \) to measure the horizon for which the sentiment score is predictive as follows:

\[
R_{t+1} ^{p_i} = \frac{P_t - P_{t-p_i}}{P_{t-p_i}}
\]

\begin{itemize}
    \item \( R_{t+1} ^{p_i} \): The return over the \( p_i \) period of the equity at time \( t+1\).
    \item \( P_t \): The value of the equity at current time \( t \).
    \item \( P_{t-p_i} \): The value of the equity at a \(p_i \) period before the current time.
\end{itemize}

On purpose, the return  \( R_{t+1} ^{p_i} \) is time stamped at time  \( t+1\) to avoid any data leakage and ensures that we have all the relevant data at the time of the computation.

\subsection{Correlation Results}
The aim is to measure the correlation between future equity market returns and the cumulative sentiment score calculated over different periods. Hence, we computed both Pearson and Spearman coefficients to evaluate the relationship between these variables two-by-two. The correlation matrices are of size 49 by 49, hence contain 2401 elements. \\

\begin{figure}[!htbp]
    \centering
    {{\includegraphics[scale=.45]{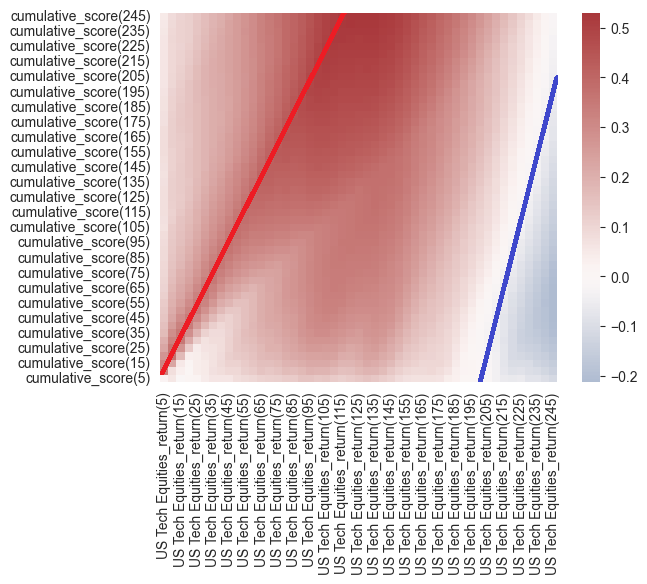} }}%
    \caption{Pearson correlation matrix of the cumulative score and the NASDAQ}
    \label{fig:corr_matrix_example}%
\end{figure}

The first experiment was to validate the difference in correlation provided by different periods for the cumulative sentiment score and forward returns. We provide in figure \ref{fig:corr_matrix_example} the result for the US Tech market.  Figures \ref{fig:US_Pearson_Index}, \ref{fig:Japan_Pearson_Index}, \ref{fig:Euro_Pearson_Index}, \ref{fig:UK_Pearson_Index}, \ref{fig:EM_Pearson_Index} provide the results for the other markets, namely US, Japan, Europe, UK and Emerging markets for the Pearson correlation matrices. Likewise, figures \ref{fig:USTech_Spearman_Index}, \ref{fig:US_Spearman_Index}, \ref{fig:Japan_Spearman_Index}, \ref{fig:Euro_Spearman_Index}, \ref{fig:UK_Spearman_Index}, \ref{fig:EM_Spearman_Index} provide the Spearmann correlation matrices for the same markets.

The overall correlation between sentiment scores and future returns is positive, as evidenced by the predominantly red color of the matrices. This positive correlation tends to increase with longer periods for both cumulative sentiment scores and forward returns, forming a diagonal pattern. However, for very long period of future returns we observe a negative correlation.

These results are consistent across markets, suggesting that the approach is robust and generalizable.

Figure \ref{fig:corr_matrix_example} showcases a red diagonal highlighting the presence of positive correlation values, while a blue diagonal signifies negative correlation patterns. In the case of positive correlation, a diagonal composed of the highest values is surrounded by other elevated values, with the values diminishing as they move away from the diagonal. We observe that the values increase for longer periods of the cumulated sentiment score. Moreover, the negative correlation pattern is evident in the long-term market return, characterized by a diagonal of non-correlated values, with a decrease of these values observed to the right of this diagonal. This pattern exists in all the other markets as proved in section \ref{sec:Robustness over the Equities Markets}.

\subsection{T-test on the correlation}
In order to validate the statistical significance of the correlation values, we applied a t-test to all the results. We focused on the p-value associated to each test. Because the number of conducted test is very large, we do all our T-test using the False Discovery Rate method.

\subsubsection{False Discovery Rate}\label{FDR}
The False Discovery Rate (FDR) is a statistical method crucial for managing the challenge of multiple comparisons in large-scale experiments, as introduced by \cite{benjamini2001control}. In contexts where numerous statistical tests are conducted simultaneously, the FDR addresses the increased risk of false positives by controlling the expected proportion of false discoveries among all significant results. This approach effectively regulates the false selection rate, ensuring that only a predetermined percentage of rejected hypotheses are likely to be false positives. The procedure is employed to rank p-values and determine a critical threshold, enabling to identify statistically significant results while managing the trade-off between sensitivity and specificity.

\subsubsection{T-Test Adaptation}
In a two-tailed t-test, the p-value signifies the probability of observing a t-statistic as extreme as the one calculated from the sample data, assuming the null hypothesis holds. For correlation values, the null hypothesis typically posits no significant correlation between the variables. The FDR adapts the threshold for improving the statistical significance assessment in a large experiment case.

\begin{figure}[!htbp]
    \centering
    {{\includegraphics[scale=0.45]{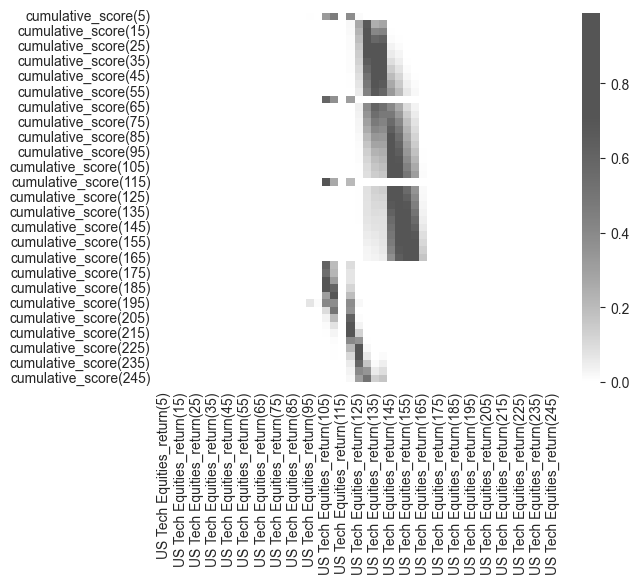} }}%
    \caption{Adjusted p-value for the Pearson correlation between the US Tech market and the cumulative sentiment score}%
    \label{fig:corr_matrix_example_NASDAQ_Pearson_cumulative_score}%
\end{figure}

In figure \ref{fig:corr_matrix_example_NASDAQ_Pearson_cumulative_score} we plot in white all the correlations whose p-value FDR adapted are below one percent and the rest, that is to say, the correlation values where we fail to reject the null hypothesis of a non significant correlation value in grey with a color scale. Most of the correlation matrix is white indicating that the correlation numbers are mostly statistically significant.  Like what we did for the correlation analysis, we can validate the tests on other equity markets. Figures \ref{fig:p_val_US_Pearson_Index}, \ref{fig:p_val_USTech_Pearson_Index}, \ref{fig:p_val_Japan_Pearson_Index}, \ref{fig:p_val_Euro_Pearson_Index}, \ref{fig:p_val_UK_Pearson_Index}, \ref{fig:p_val_EM_Pearson_Index} show that all equitites markets exhibit similar behavior for the p-values of Pearson correlation while figures \ref{fig:p_val_US_Spearman_Index}, \ref{fig:p_val_USTech_Spearman_Index}, \ref{fig:p_val_Japan_Spearman_Index}, \ref{fig:p_val_Euro_Pearson_Index}, \ref{fig:p_val_UK_Pearson_Index}, \ref{fig:p_val_EM_Spearman_Index} show that all equitites markets exhibit similar behavior for the p-values of Spearman correlation

\subsubsection{The Mitigated Matrix}
Consideration should be given exclusively to correlation values demonstrating statistical significance. Our aim is to adjust each correlation in accordance with its corresponding p-value. As illustrated in Figure \ref{fig:corr_matrix_example_NASDAQ_Pearson_cumulative_score}, the correlation matrix is modified using a gradient approach. Specifically, correlation values are retained as-is when p-values suggest statistical significance. Conversely, in instances of increasing p-value, the correlations are adjusted as follows:

\begin{equation}
\rho_{i,j}^{\text{mitigated}} = \rho_{i,j} \times (1 - p_{i,j})
\end{equation}

\begin{figure}[!htbp]
    \centering
    {{\includegraphics[scale=0.45]{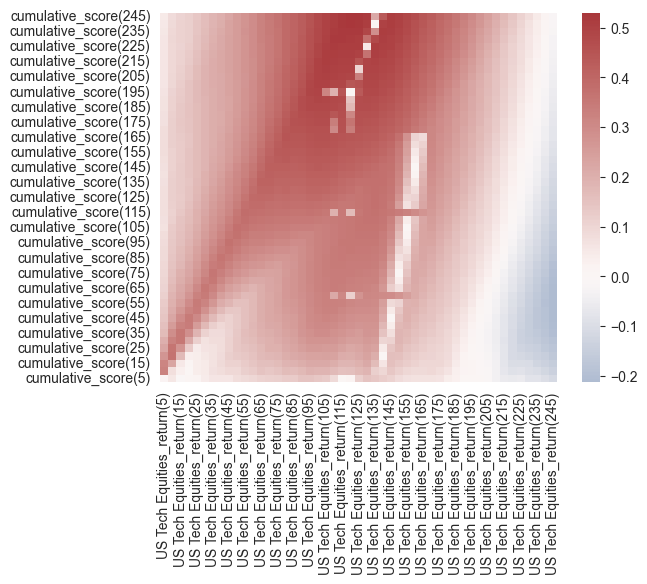} }}%
    \caption{Mitigated correlation between the US Tech market and the cumulative sentiment score}%
    \label{fig:mitigated_ustech_correlation}%
\end{figure}

Here, $\rho_{i,j}$ represents the correlation coefficient, and $p_{i,j}$ denotes the associated p-value. Figure \ref{fig:mitigated_ustech_correlation} displays the resulting mitigated correlation matrix. This method allows for the prioritization of statistically significant correlations without excessive discrimination.

Analysis reveals that the matrix's region of interest is predominantly significant. Non-significant values are found in longer horizons for cumulative\_score and Equity return. These findings are applicable across various equity markets for both Pearson and Spearman correlations. Figure \ref{fig:USTech_Pearson_mitig}, \ref{fig:US_Pearson_mitig}, \ref{fig:Japan_Pearson_mitig}, \ref{fig:UK_Pearson_mitig}, \ref{fig:Euro_Pearson_mitig}, \ref{fig:EM_Pearson_mitig},
\ref{fig:USTech_Spearman_mitig}, \ref{fig:US_Spearman_mitig}, \ref{fig:Japan_Spearman_mitig}, \ref{fig:UK_Spearman_mitig}, \ref{fig:Euro_Spearman_mitig}, \ref{fig:EM_Spearman_mitig}  corroborate these results.

\subsubsection{The Short Term Correlation}
The correlation exhibits notable values within the range of $[-0.30, 0.53]$. To emphasize statistical significance, we focus exclusively on the matrix section where p-values are below $0.01$, representing the white area. This selection yields significant correlation results for the mid-term return and cumulative sentiment score lag.

\subsubsection{The Best Combinations}

Among all the coefficients, we can exclude the non significant ones according to a t-test, hence with p-value exceeding the $0.01$ threshold. We could obtain the duo of variables that obtain the highest and lowest correlation values in table \ref{tab:sentiment_pearson_highest_corr_by_equities} and \ref{tab:sentiment_highest_pearson_neg_corr_by_equities} respectively.

\begin{table}[!htbp]
    \centering
    \resizebox{\columnwidth}{!}{%
    \begin{tabular}{|c|c|c|}
        \hline
        \textbf{Score} & \textbf{Equity} & \textbf{Positive Correlation} \\
        \hline
        $S_{245}$ & US Tech(125) & 0.53 \\
        \hline
        $S_{205}$ & US(95) & 0.47 \\
        \hline
        $S_{245}$ & Japan(135) & 0.27 \\
        \hline
        $S_{80}$ & Europe(35) & 0.22 \\
        \hline
        $S_{80}$ & UK(35) & 0.25 \\
        \hline
        $S_{245}$ & Emerging(125) & 0.43 \\
        \hline
    \end{tabular}
    }
    \caption{Highest Pearson positive correlation values by equities}
    \label{tab:sentiment_pearson_highest_corr_by_equities}
\end{table}

\begin{table}[!htbp]
    \centering
    \resizebox{\columnwidth}{!}{%
    \begin{tabular}{|c|c|c|}
        \hline
        \textbf{Score} & \textbf{Equity} & \textbf{Negative Correlation} \\
        \hline
        $S_{245}$ & US Tech(245) & -0.26 \\
        \hline
        $S_{5}$ & US(245) & -0.31 \\
        \hline
        $S_{245}$ & Japan(245) & -0.19 \\
        \hline
        $S_{5}$ & Europe(245) & -0.30 \\
        \hline
        $S_{5}$ & UK(210) & -0.18 \\
        \hline
        $S_{245}$ & Emerging(245) & -0.14 \\
        \hline
    \end{tabular}
    }
    \caption{Highest Pearson negative correlation values by equities}
    \label{tab:sentiment_highest_pearson_neg_corr_by_equities}
\end{table}

We remind that $S_{d}$ represents the cumulative score denoted in the matrix as "cumulative\_score(d)".\\ 

The analysis reveals a clear, positive relationship between the cumulative score and equity returns, with the strength of the correlation intensifying as the lag size of the cumulative score increases. Interestingly, as we delve into deeper cumulative scores, the negative correlation diminishes. There is a discernible trade-off concerning the lag of the cumulative score: seeking an optimal balance is crucial, as the cumulative score lags behind the equity market. We aim to maximize the correlation while maintaining a current score reflective of the market's status. For instance, opting for a substantial lag in the cumulative score may yield a strong correlation, yet the estimator's time relevance could be compromised. This dynamic is evident in the correlation matrix, where red signifies positive correlation and blue indicates negative correlation, guiding us towards a precise analysis. Markets demonstrate different degrees of sensitivity to the timing of news, with the cumulative score's correlation extending over a more extended period than previously observed with sentiment scores. The investigation into the relationship between cumulative scores and equity returns illuminates the crucial dynamics of lag impact. The subsequent section will delve into the intricate trade-off that exists between the lag value of the score and the intensity of the signal it provides.

\section{Trade-Off Analysis of Financial Indicators}\label{sec:Trade-Off Analysis of Financial Indicators}

The investigation into the relationship between cumulative scores \( S_d \) and equity returns unveils the pivotal dynamics of market reaction delays. The forthcoming analysis explores the nuanced trade-off between the depth of the cumulative score—reflected by the subscript \( d \) in \( S_d \)—and the predictive signal's intensity it conveys. The term \( d \) represents the depth of analysis, encapsulating the cumulative effect of sentiment over a defined period.

The depth of the cumulative score, denoted as \( S_d \), is mathematically defined as the aggregate sentiment measured over a period \( d \). This period reflects the span over which the sentiment data is cumulated, not to be confused with the delay in market reaction. The delay in market impact is instead associated with the temporal shift applied to the equity return data, which is examined against the cumulative sentiment scores.

The correlation value, represented by \( \rho \), quantifies the strength and direction of the linear relationship between the financial indicator's cumulative score \( S_d \) and the shifted equity returns. The mean correlation value for different prediction horizons, ranging from 1 to 12 months, is computed as follows:

\begin{equation}
\bar{\rho}_{\text{horizon}}(i) = \frac{1}{s\times(j+1)}\sum_{k=1}^{s\times(j+1)} \rho_{i,k}
\end{equation}

where \( \bar{\rho}_{\text{horizon}}(i) \) is the mean correlation at the \( i^{th} \) cumulative value for a given horizon, and \( \rho_{i,k} \) is the correlation value at the \( i^{th} \) cumulative value for the \( k^{th} \) shifted time point within the horizon. The term \( s\times(j+1) \) denotes the number of discrete time intervals encapsulated within the horizon, where \( j \in \{0, 1, \ldots, 11\} \) and $s$ is the number of equity return included for mean computation.

\begin{figure}[!htbp]
  \centering
  \includegraphics[scale=.5]{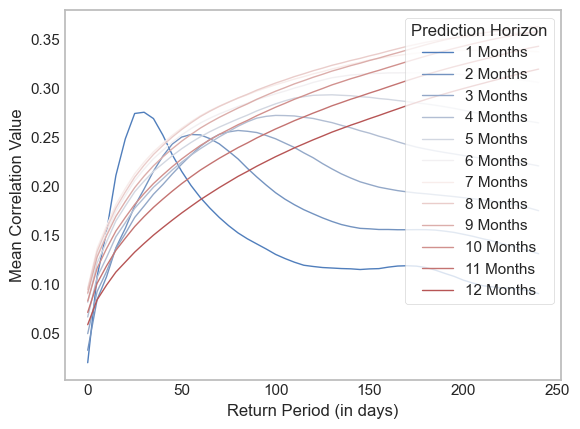}
  \caption{US Tech Equity: Mean correlation of cumulative score against shifted returns across horizons}
  \label{fig:tradeoffgraph}
\end{figure}

The objective of the analysis is to determine the optimal depth \( d_{\text{opt}} \) of \( S_d \) that maximizes \( \rho \), while still being timely enough to provide practical predictive utility for market reactions. This optimal point is characterized by the highest mean correlation value that can be achieved before the utility of the cumulative score is compromised by its stale reflection of market sentiment. Like what we did for the correlation analysis, we can perform the same analysis for the other equity markets. This is provided by figures \ref{fig:tradeoff_USTech}, \ref{fig:tradeoff_US}, \ref{fig:tradeoff_Japan}, \ref{fig:tradeoff_Euro}, \ref{fig:tradeoff_UK}, \ref{fig:tradeoff_EM}.

\subsection{Optimal Point Determination}
The apex of the curve in Figure \ref{fig:tradeoffgraph} indicates the optimal depth \( d_{\text{opt}} \) of \( S_d \), at which the mean correlation \( \bar{\rho} \) is maximized. This peak represents the ideal balance between comprehensive sentiment analysis and timely market prediction, ensuring the cumulative score's relevance and predictive power.

To ascertain \( d_{\text{opt}} \), we locate the curve's highest point, which signifies the strongest linear relationship between \( S_d \) and market performance, without undue delay. Table \ref{tab:equity_rho_values} provides the optimal period for each equity market over a prediction horizon of one month. The step size $s$ is $4$ and includes the market return on 20 days. 

\begin{table}[!htbp]
\centering
\begin{tabular}{|c|c|}
\hline
\textbf{Equity} & \textbf{\( d_{\text{opt}} \)} \\\hline
US Tech & 40 \\\hline
US & 30 \\\hline
Japan & 40 \\\hline
EU & 25 \\\hline
UK & 30 \\\hline
EM & 30 \\
\hline
\end{tabular}
\caption{Mean correlation values for different equities over one month}
\label{tab:equity_rho_values}
\end{table}

\section{Robustness over the Equities Markets}\label{sec:Robustness over the Equities Markets}

This section examines the robustness of the identified pattern across different equity markets. The question arises whether a universal pattern exists within these markets. To address this, we compare each matrix with the average correlation matrix, representing the common pattern, and assess the distance in terms of standard deviation of each matrix from this common pattern. Like for the rest of the paper on other indicators, we can notice consistency across equities markets.

\subsection{Computation of Mean Matrix and Standard Deviation}
The mean matrix, denoted as \( Z \), is computed as the average of all correlation matrices:
\begin{equation}
Z =  \frac{1}{n}  \left( \sum_{k=1}^{n} m_{i,j}^k \right)  
\label{eq:mean_matrix}
\end{equation}
     
where \(\left( m_{i,j}^k \right) \) represents the $i,j$ correlation matrix coefficient of the $k$ market and \( n \) is the total number of markets. Strictly speaking, the mean matrix is computed for each cell as the mean across all markets. Likewise, for each matrix cell, we compute the standard deviation of correlations across all markets
\[ \Sigma(Z) = \frac{1}{\sqrt{n-1} }   \left( \sum_{k=1}^{n} \left( m_{i,j}^k - z_{i,j} \right)^2 \right)  \]
where  \(  z_{i,j}  = \sum_{k=1}^{n} m_{i,j}^k  / n  \) is the coefficient of the mean matrix presented in equation \ref{eq:mean_matrix}

\subsection{Element-wise T-test Analysis}
In order to ensure proprer resizing of each correlation as well as the average correlation matrix, we first z-score them as follows:
\begin{equation}
   \Tilde{M}^k_{ij} = \frac{ M_{ij}^k - \bar{M}^k_{ij}}{\Sigma(M)_{ij}} 
\end{equation}

For each market matrix with upper index $k$, we conduct an element-wise T-test comparing it to the mean matrix \( Z \). The T-statistic is computed elementwise as:
\begin{equation}
   T_{ij} = \frac{ M_{ij}^k -  Z_{ij}}{\Sigma(Z)_{ij}} 
\end{equation}

The p-values are computed using two-tails test:
\begin{equation}
    p = 1 - 2 \times (1 - \text{CDF}_{student}(\lvert T \rvert)) 
\end{equation}



\subsection{Analysis of P-Value Results}
Table \ref{table:p-values_by_equities_market} presents the percentage of each equity market matrix where the p-value falls below the 0.01 significance threshold:

\begin{table}[!h]
\centering
\begin{tabular}{|l|c|}
\hline
\textbf{Equity Market} & \textbf{\% of Matrix} \\ \hline
US Tech               & 80 \\ \hline
US                    & 91 \\ \hline
Japan                 & 92 \\ \hline
Euro                  & 86 \\ \hline
United Kingdom        & 55 \\ \hline
Emerging              & 75 \\ \hline
\end{tabular}
\caption{Proportion of Each Equity Matrix Validating the Common Pattern}
\label{table:p-values_by_equities_market}
\end{table}

A score of 100\% implies that the matrix perfectly follows the pattern of the mean matrix, while a score of 0\% indicates no common pattern with the mean matrix.

The results indicate a significant presence of the identified pattern across all markets, with an especially pronounced effect in the Japanese market (99\%). The US Technology and US General markets exhibit substantial percentages (78\% and 69\% respectively). This variation suggests a differential impact of sentiment scores on equity returns across these markets.

The high percentages in the Euro, United Kingdom, and Emerging Markets (ranging from 84\% to 94\%) further reinforce the ubiquity of the pattern. These findings collectively suggest that sentiment scores consistently influence equity returns across diverse global markets, underpinning the robustness of the identified pattern.

This analysis confirms the existence of a common pattern across various equity markets, linking sentiment scores to equity returns. The consistency of significant p-values across markets underscores the widespread impact of investor sentiment on market movements, presenting valuable insights for market analysis and investment strategies.

\subsection{Matrix quantile distance}
A second method consists in doing a quantile difference test between each market correlation matrix and the average over each market. Although this approach is less well-known than the standard correlation t-test, converting correlation matrices into quantiles for each cell and then computing their average absolute difference to judge the quantile distance is a method to judge if two matrices share a similar profile. This approach makes sense for several reasons:

\begin{itemize}
  \item \textbf{Robustness:} Quantiles are less affected by outliers compared to raw correlation values. This can give a more robust comparison, especially in the presence of extreme values.
  \item \textbf{Normalization:} It normalizes the scale of comparison. Since correlation coefficients are bounded between -1 and 1, converting them to quantiles puts them on a uniform scale.
  \item \textbf{Sensitivity to Distribution:} This method is sensitive to the distribution of correlation coefficients across the matrices. By using quantiles, you're comparing the relative positions of correlation coefficients, which can be more informative about the similarity in patterns of correlation.
  \item \textbf{Interpretable Metric:} The average absolute difference is an easily interpretable metric that quantifies the average discrepancy between the matrices in terms of their quantile-transformed correlations.
\end{itemize}

Mathematically, if \( C_1 \) and \( C_2 \) are two correlation matrices, converting them to quantiles involves replacing each correlation coefficient with its corresponding quantile rank within the matrix that we denote for each matrix \(i,j\) cell as \(Q_{ij}^1 \) and \(Q_{ij}^2 \) respectively.  The average absolute difference is calculated as \( \frac{1}{n^2} \sum_{i=1}^{n}\sum_{j=1}^{n} |Q_{ij}^1  - Q_{ij}^2| \), where \( n \) is the dimension of the matrices. This value gives an overall measure of how different the two matrices are in their correlation structure. 

Table \ref{table:quantile_distance} displays the proportion of each equity market matrix with quantile difference above ten percents. The results exhibit consistency with the previous method table \ref{table:p-values_by_equities_market}, confirming us that the sentiment news is consistent accross major equities markets.

\begin{table}[!htbp]
\centering
\begin{tabular}{|l|c|}
\hline
\textbf{Equity Market} & \textbf{\% of Matrix} \\ \hline
US Tech               & 80 \\ \hline
US                    & 99 \\ \hline
Japan                 & 92 \\ \hline
Euro                  & 89 \\ \hline
United Kingdom        & 51 \\ \hline
Emerging              & 72 \\ \hline
\end{tabular}
\caption{Proportion of Each Equity Matrix Validating the Common Pattern using quantile distance over 10\%}
\label{table:quantile_distance}
\end{table}

\section{Conclusion}\label{sec:Conclusion}
In this paper, we look at the equity market reaction to market news sentiment. We document significant correlations between news market sentiment and equity returns regarding the cumulative sentiment score. We also show that the correlation reverts to a negative correlation over longer horizons. We validate that this behavior exists in other equity markets, validating the robustness of the pattern. We suggest an optimal period that balances the trade-off between the market's reactivity to new information and the strength of correlation between sentiment score and forward equities returns. \\
Future research could elaborate on this sentiment score to suggest a systematic NLP based long short strategy on world wide equity indices.

\clearpage
\twocolumn
\section{Bibliographical References}
\bibliographystyle{ACM-Reference-Format}
\bibliography{main}

\clearpage

\onecolumn
\appendix

\section{Appendix}

\subsection{Cumulative Sentiment Score}

\subsubsection{Pearson Correlation Results}

\begin{figure}[!htbp]
    \centering
    {{\includegraphics[scale=0.65]{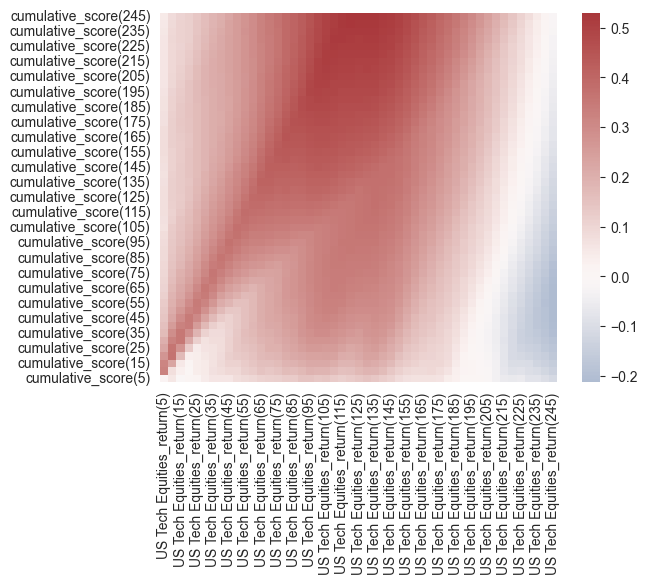} }}%
    \caption{Pearson correlation between the USTech and the cumulative sentiment score}\label{fig:USTech_Pearson_Index}%
\end{figure}

\begin{figure}[!htbp]
    \centering
    {{\includegraphics[scale=0.65]{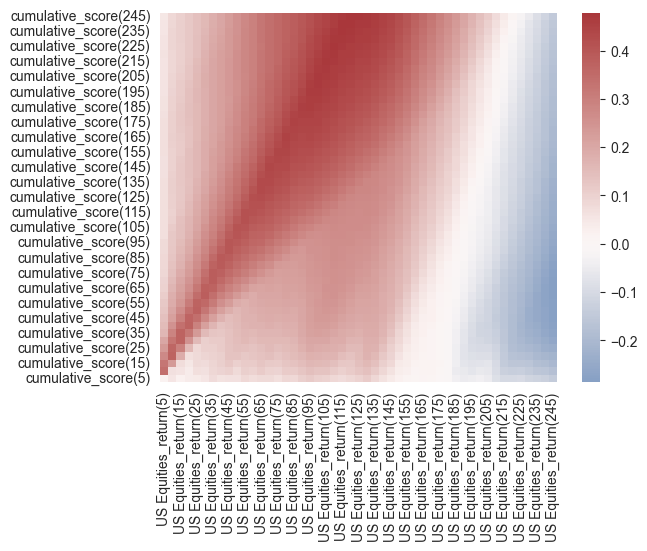} }}%
    \caption{Pearson correlation between the US and the cumulative sentiment score}\label{fig:US_Pearson_Index}%
\end{figure}

\begin{figure}[!htbp]
    \centering
    {{\includegraphics[scale=0.65]{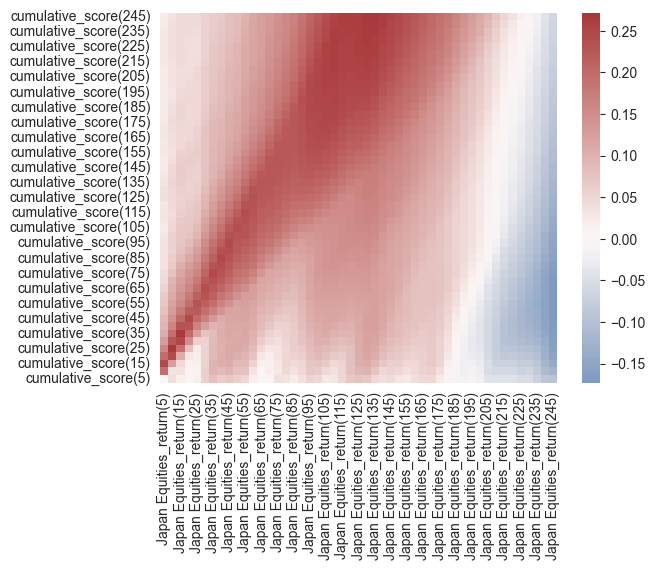} }}%
    \caption{Pearson correlation between Japan and the cumulative sentiment score}\label{fig:Japan_Pearson_Index}%
\end{figure}

\begin{figure}[!htbp]
    \centering
    {{\includegraphics[scale=0.65]{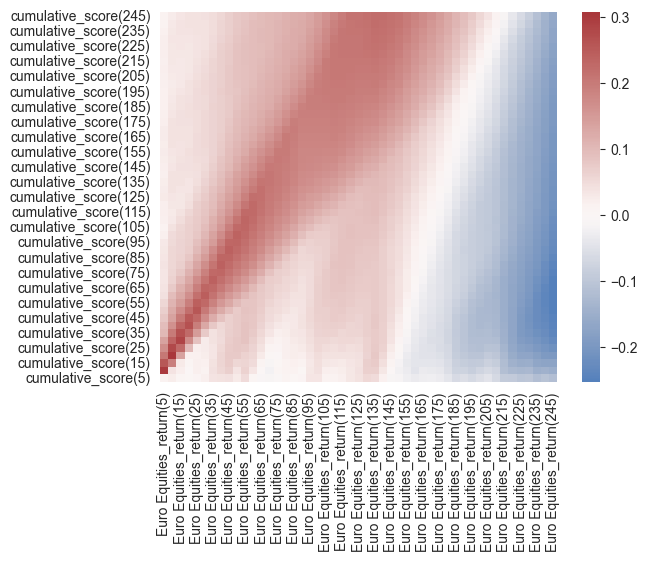} }}%
    \caption{Pearson correlation between Euro and the cumulative sentiment score}\label{fig:Euro_Pearson_Index}%
\end{figure}

\begin{figure}[!htbp]
    \centering
    {{\includegraphics[scale=0.65]{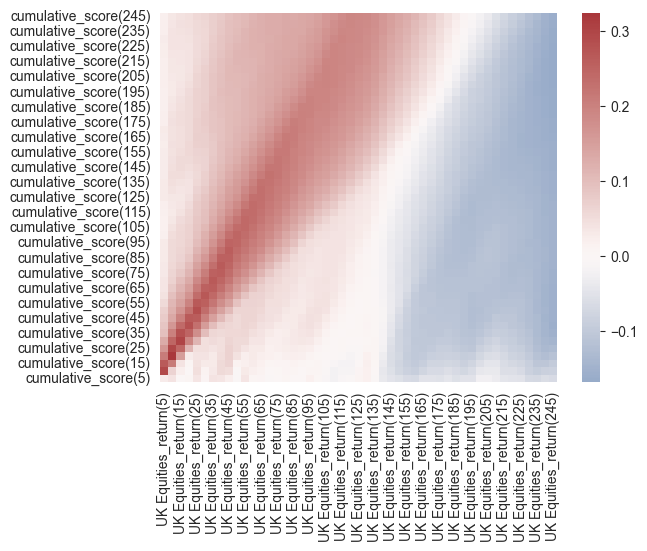} }}%
    \caption{Pearson correlation between the UK and the cumulative sentiment score}\label{fig:UK_Pearson_Index}%
\end{figure}

\begin{figure}[!htbp]
    \centering
    {{\includegraphics[scale=0.65]{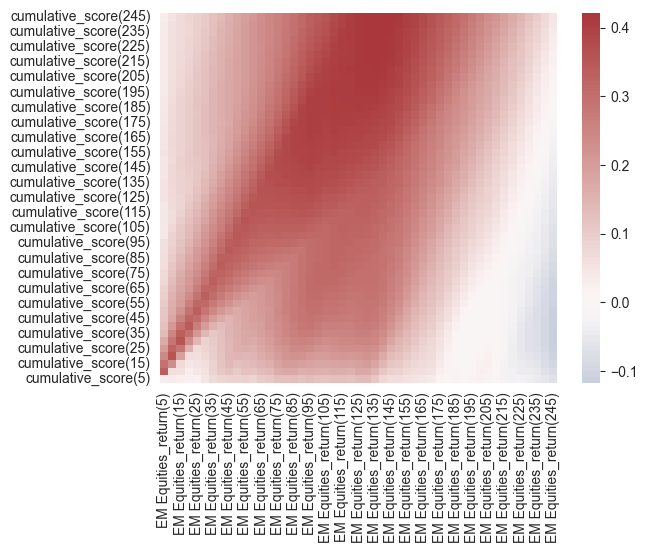} }}%
    \caption{Pearson correlation between EM and the cumulative sentiment score}\label{fig:EM_Pearson_Index}%
\end{figure}
\clearpage

\subsubsection{P-value Pearson Correlation Results}

\begin{figure}[!htbp]
    \centering
    {{\includegraphics[scale=0.65]{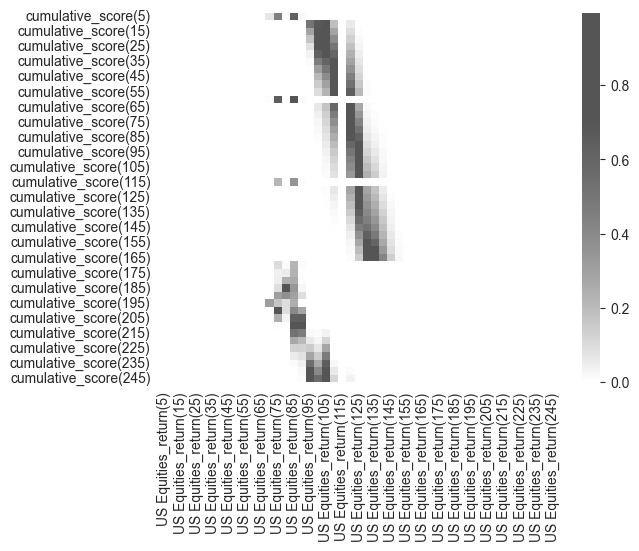} }}%
    \caption{P-value for Pearson correlation between the U.S. and the cumulative sentiment score}\label{fig:p_val_US_Pearson_Index}%
\end{figure}

\begin{figure}[!htbp]
    \centering
    {{\includegraphics[scale=0.65]{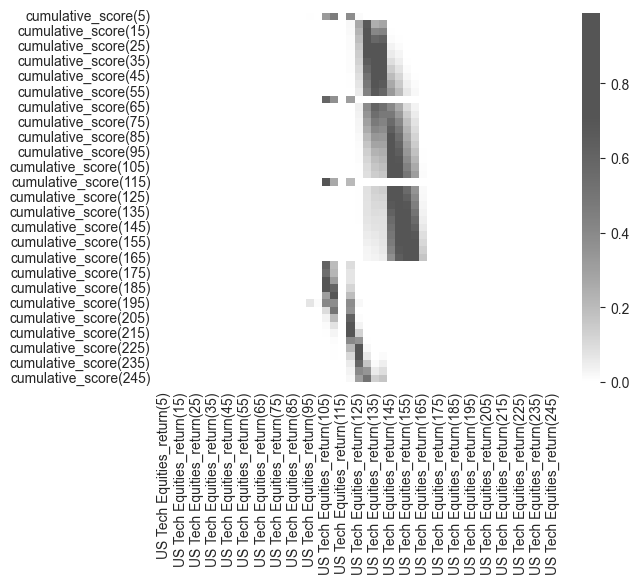} }}%
    \caption{P-value for Pearson correlation between USTech and the cumulative sentiment score}\label{fig:p_val_USTech_Pearson_Index}%
\end{figure}

\begin{figure}[!htbp]
    \centering
    {{\includegraphics[scale=0.65]{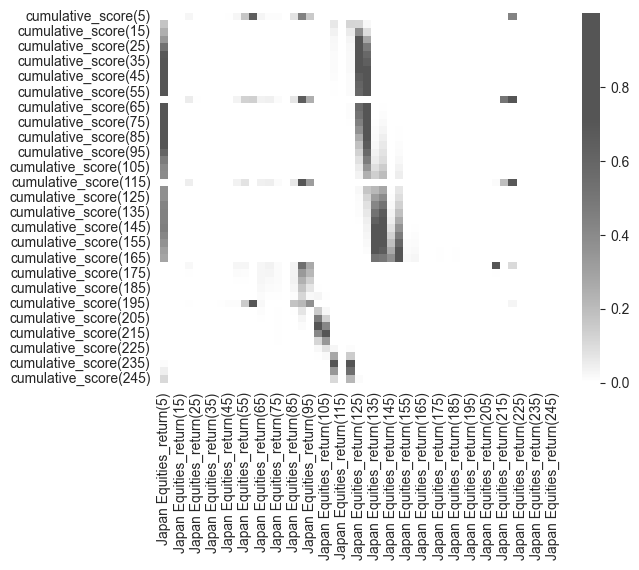} }}%
    \caption{P-value for Pearson correlation between Japan and the cumulative sentiment score}\label{fig:p_val_Japan_Pearson_Index}%
\end{figure}

\begin{figure}[!htbp]
    \centering
    {{\includegraphics[scale=0.65]{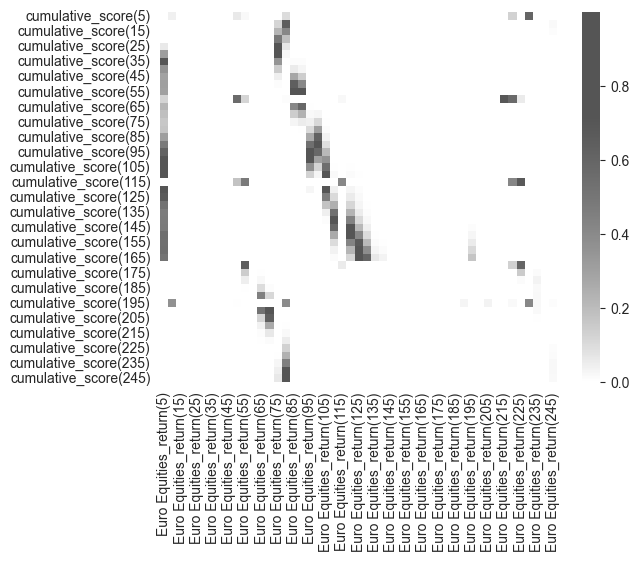} }}%
    \caption{P-value for Pearson correlation between Euro and the cumulative sentiment score}\label{fig:p_val_Euro_Pearson_Index}%
\end{figure}

\begin{figure}[!htbp]
    \centering
    {{\includegraphics[scale=0.65]{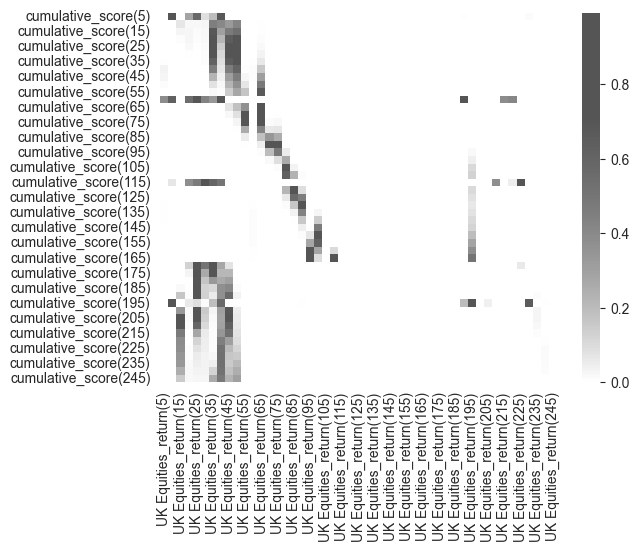} }}%
    \caption{P-value for Pearson correlation between the UK and the cumulative sentiment score}\label{fig:p_val_UK_Pearson_Index}%
\end{figure}

\begin{figure}[!htbp]
    \centering
    {{\includegraphics[scale=0.65]{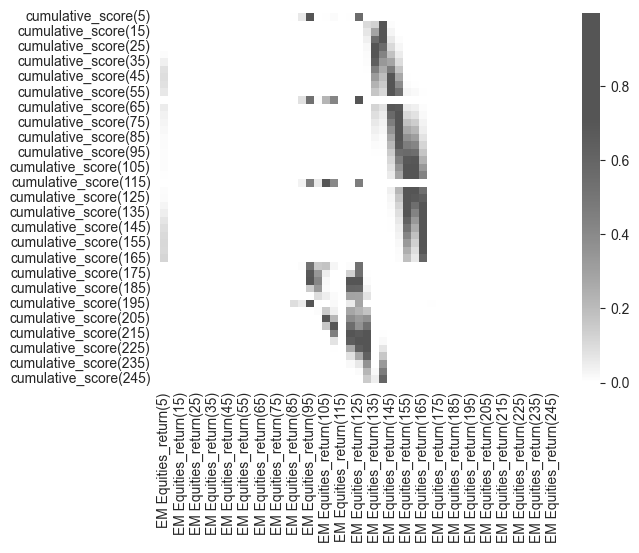} }}%
    \caption{P-value for Pearson correlation between EM and the cumulative sentiment score}\label{fig:p_val_EM_Pearson_Index}%
\end{figure}
\clearpage

\subsubsection{Mitigated Pearson Correlation Results}

\begin{figure}[!htbp]
    \centering
    {{\includegraphics[scale=0.65]{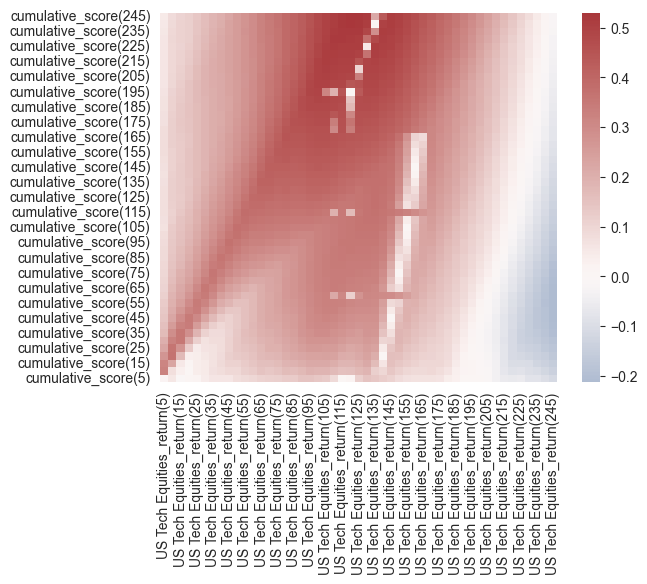} }}%
    \caption{Mitigated Pearson correlation between the USTech and the cumulative sentiment score}\label{fig:USTech_Pearson_mitig}%
\end{figure}

\begin{figure}[!htbp]
    \centering
    {{\includegraphics[scale=0.65]{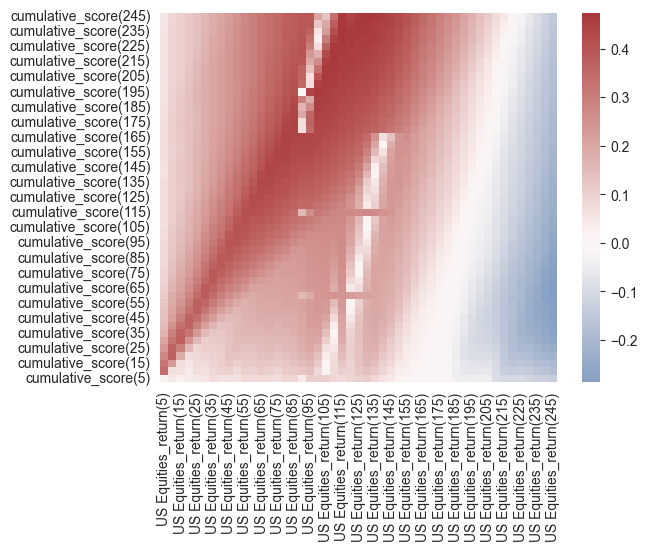} }}%
    \caption{Mitigated Pearson correlation between the US and the cumulative sentiment score}\label{fig:US_Pearson_mitig}%
\end{figure}

\begin{figure}[!htbp]
    \centering
    {{\includegraphics[scale=0.65]{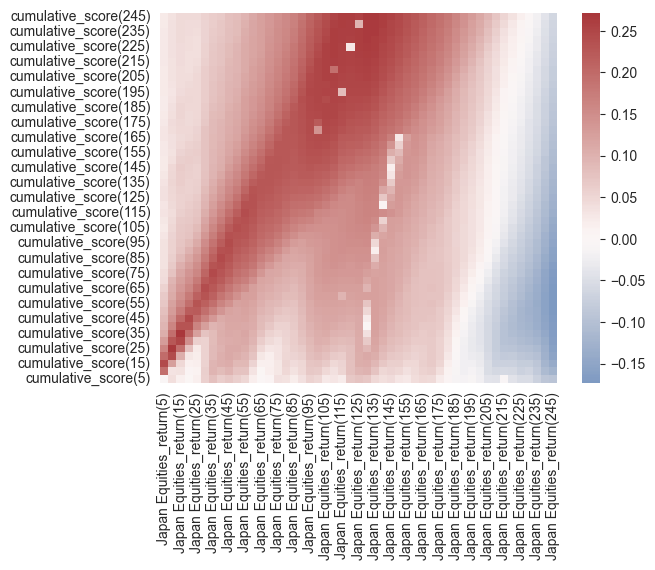} }}%
    \caption{Mitigated Pearson correlation between Japan and the cumulative sentiment score}\label{fig:Japan_Pearson_mitig}%
\end{figure}

\begin{figure}[!htbp]
    \centering
    {{\includegraphics[scale=0.65]{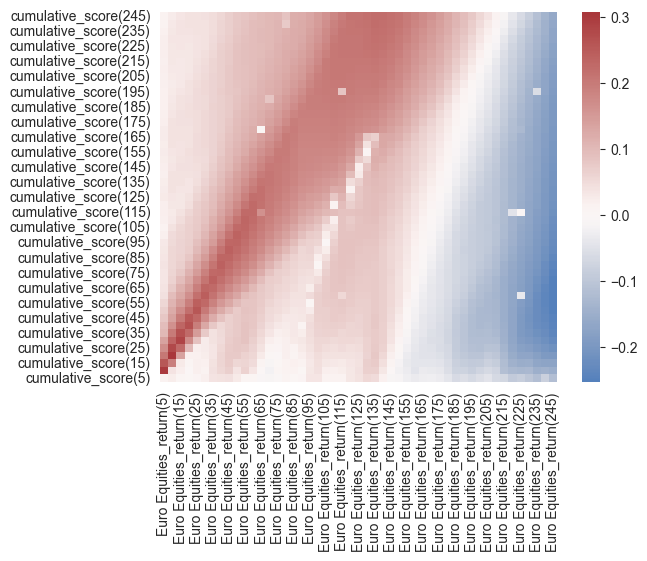} }}%
    \caption{Mitigated Pearson correlation between Euro and the cumulative sentiment score}\label{fig:Euro_Pearson_mitig}%
\end{figure}

\begin{figure}[!htbp]
    \centering
    {{\includegraphics[scale=0.65]{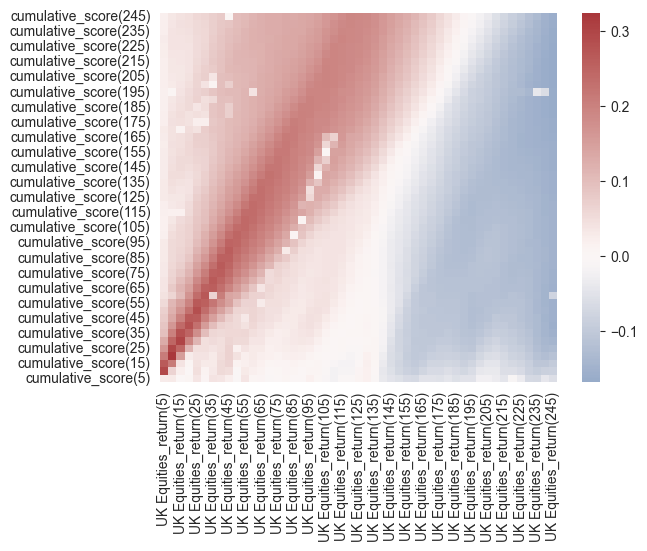} }}%
    \caption{Mitigated Pearson correlation between the UK and the cumulative sentiment score}\label{fig:UK_Pearson_mitig}%
\end{figure}

\begin{figure}[!htbp]
    \centering
    {{\includegraphics[scale=0.65]{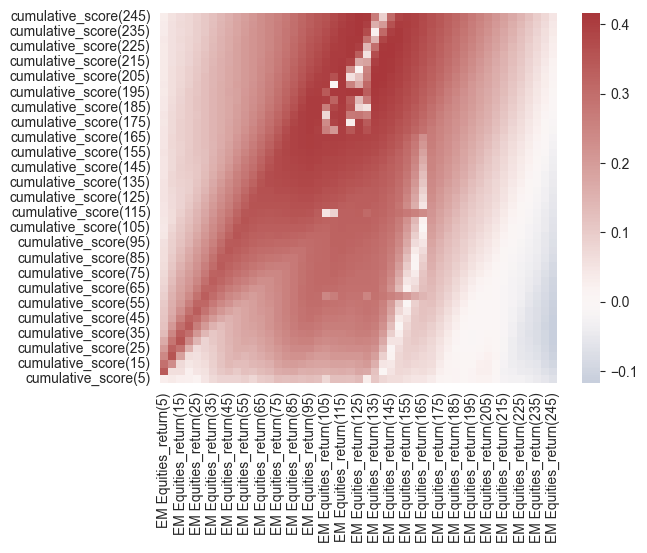} }}%
    \caption{Mitigated Pearson correlation between EM and the cumulative sentiment score}\label{fig:EM_Pearson_mitig}%
\end{figure}
\clearpage

\subsubsection{Spearman Correlation Results}

\begin{figure}[!htbp]
    \centering
    {{\includegraphics[scale=0.65]{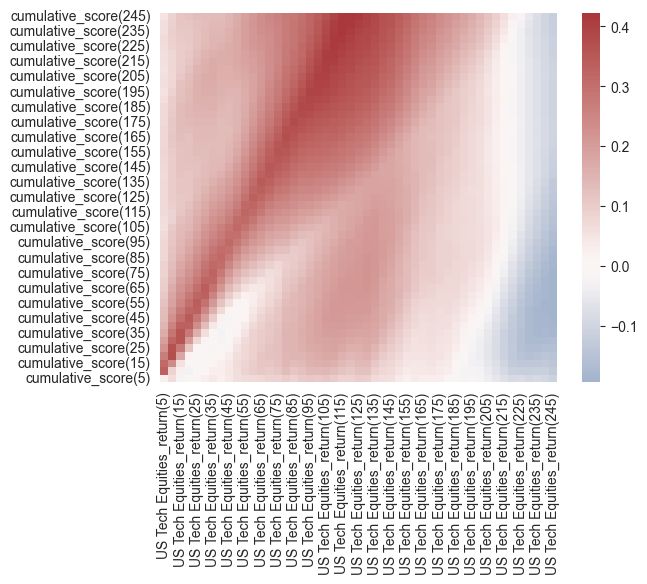} }}%
    \caption{Spearman correlation between USTech and the cumulative sentiment score}\label{fig:USTech_Spearman_Index}%
\end{figure}

\begin{figure}[!htbp]
    \centering
    {{\includegraphics[scale=0.65]{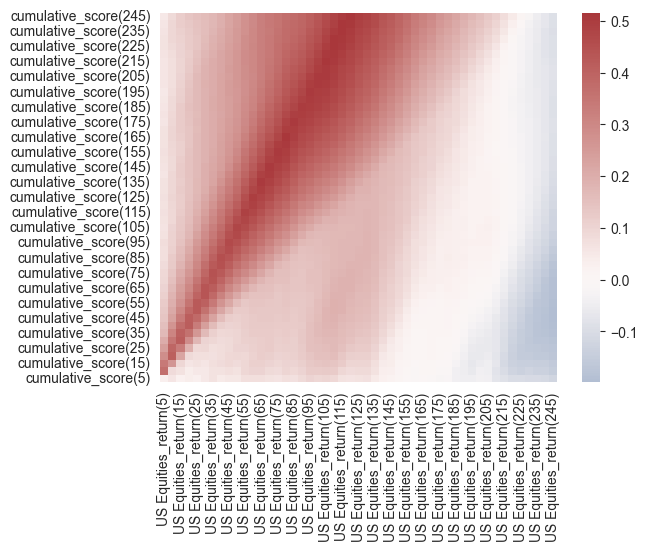} }}%
    \caption{Spearman correlation between the U.S. and the cumulative sentiment score}\label{fig:US_Spearman_Index}%
\end{figure}

\begin{figure}[!htbp]
    \centering
    {{\includegraphics[scale=0.65]{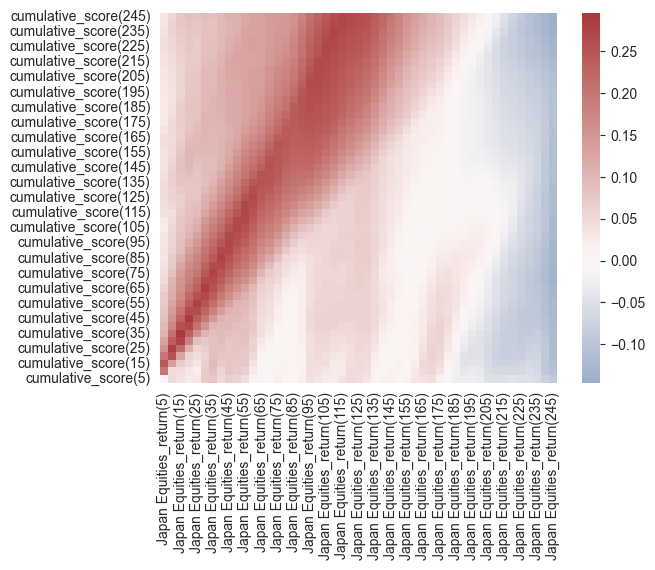} }}%
    \caption{Spearman correlation between Japan and the cumulative sentiment score}\label{fig:Japan_Spearman_Index}%
\end{figure}

\begin{figure}[!htbp]
    \centering
    {{\includegraphics[scale=0.65]{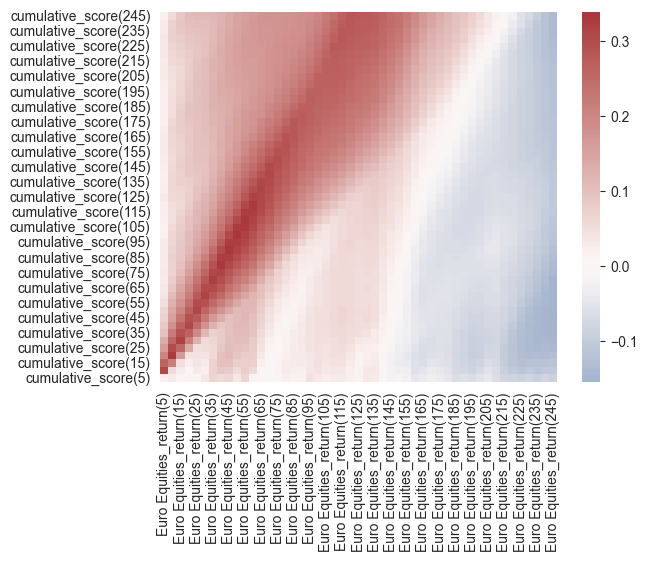} }}%
    \caption{Spearman correlation between Euro and the cumulative sentiment score}\label{fig:Euro_Spearman_Index}%
\end{figure}

\begin{figure}[!htbp]
    \centering
    {{\includegraphics[scale=0.65]{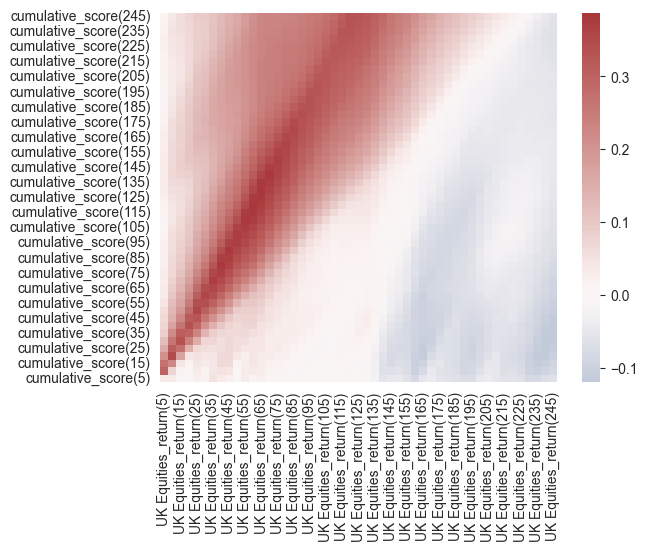} }}%
    \caption{Spearman correlation between the UK and the cumulative sentiment score}\label{fig:UK_Spearman_Index}%
\end{figure}

\begin{figure}[!htbp]
    \centering
    {{\includegraphics[scale=0.65]{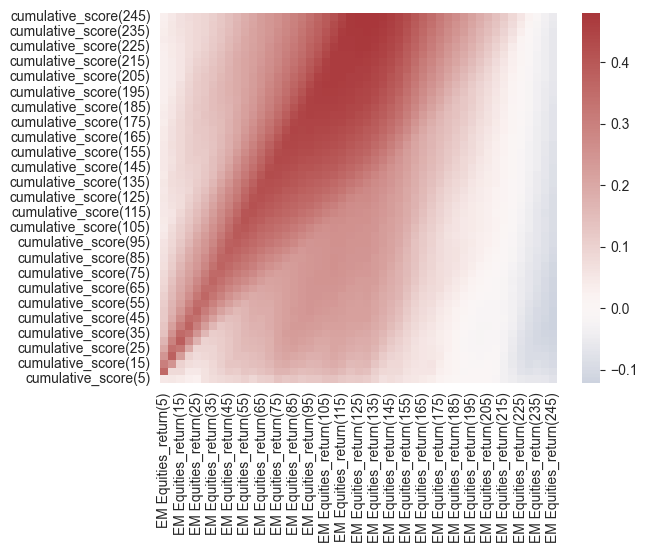} }}%
    \caption{Spearman correlation between EM and the cumulative sentiment score}\label{fig:EM_Spearman_Index}%
\end{figure}
\clearpage

\subsubsection{P-value Spearman Correlation Results}
\begin{figure}[!htbp]
    \centering
    {{\includegraphics[scale=0.65]{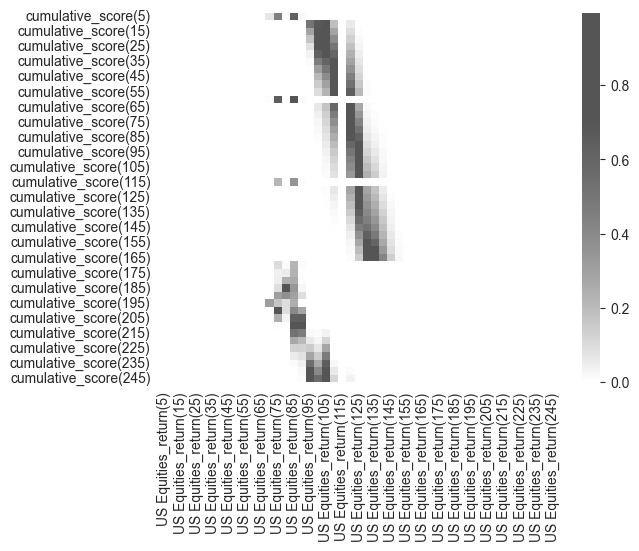} }}%
    \caption{P-value for Spearman correlation between the U.S. and the cumulative sentiment score}\label{fig:p_val_US_Spearman_Index}%
\end{figure}

\begin{figure}[!htbp]
    \centering
    {{\includegraphics[scale=0.65]{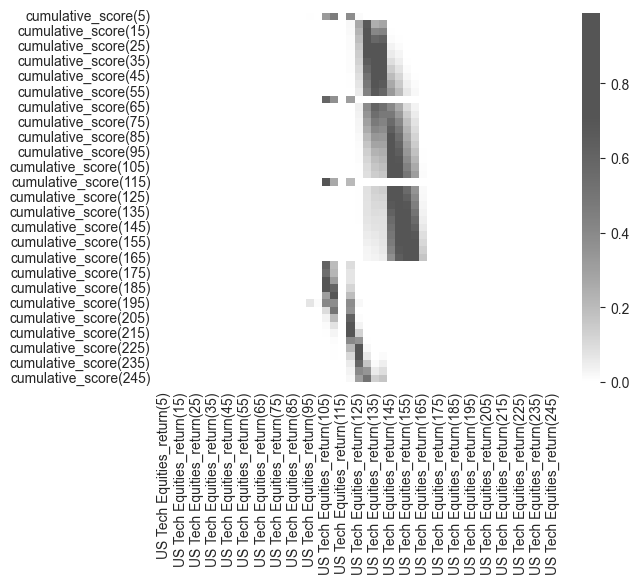} }}%
    \caption{P-value for Spearman correlation between USTech and the cumulative sentiment score}\label{fig:p_val_USTech_Spearman_Index}%
\end{figure}

\begin{figure}[!htbp]
    \centering
    {{\includegraphics[scale=0.65]{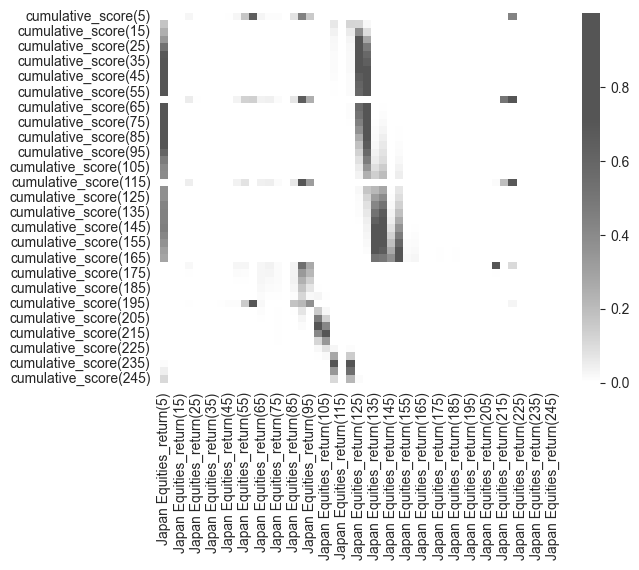} }}%
    \caption{P-value for Spearman correlation between Japan and the cumulative sentiment score}\label{fig:p_val_Japan_Spearman_Index}%
\end{figure}

\begin{figure}[!htbp]
    \centering
    {{\includegraphics[scale=0.65]{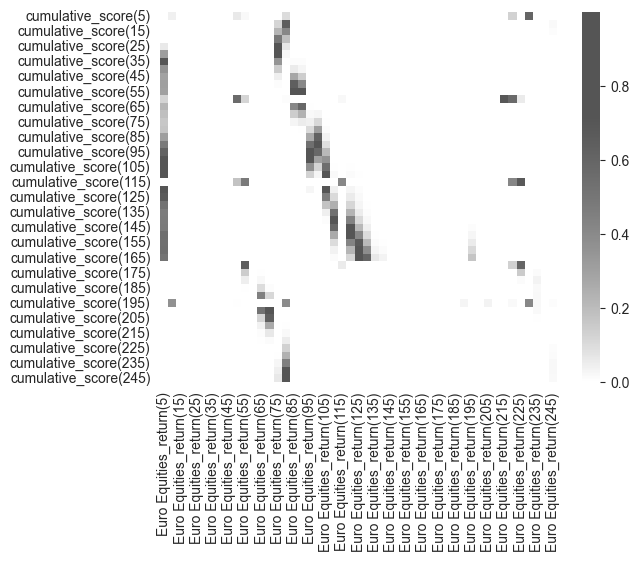} }}%
    \caption{P-value for Spearman correlation between Euro and the cumulative sentiment score}\label{fig:p_val_Euro_Spearman_Index}%
\end{figure}

\begin{figure}[!htbp]
    \centering
    {{\includegraphics[scale=0.65]{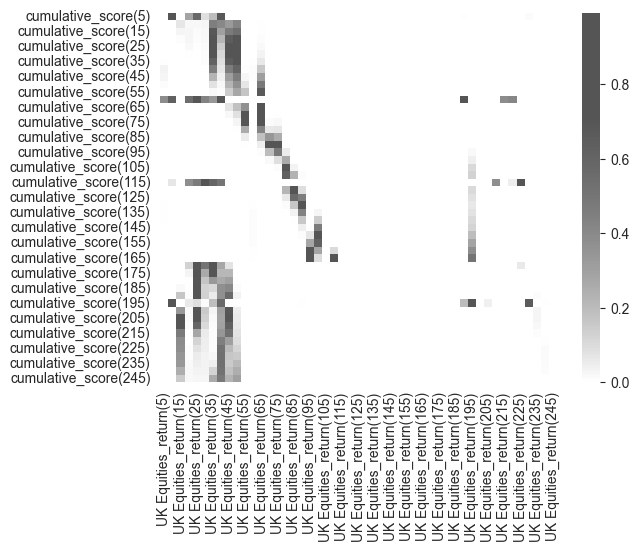} }}%
    \caption{P-value for Spearman correlation between the UK and the cumulative sentiment score}\label{fig:p_val_UK_Spearman_Index}%
\end{figure}

\begin{figure}[!htbp]
    \centering
    {{\includegraphics[scale=0.65]{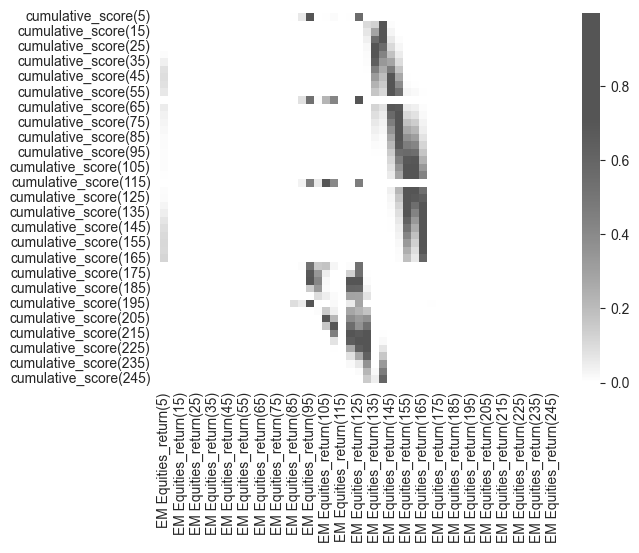} }}%
    \caption{P-value for Spearman correlation between EM and the cumulative sentiment score}\label{fig:p_val_EM_Spearman_Index}%
\end{figure}
\clearpage

\subsubsection{Mitigated Spearman Correlation Results}

\begin{figure}[!htbp]
    \centering
    {{\includegraphics[scale=0.65]{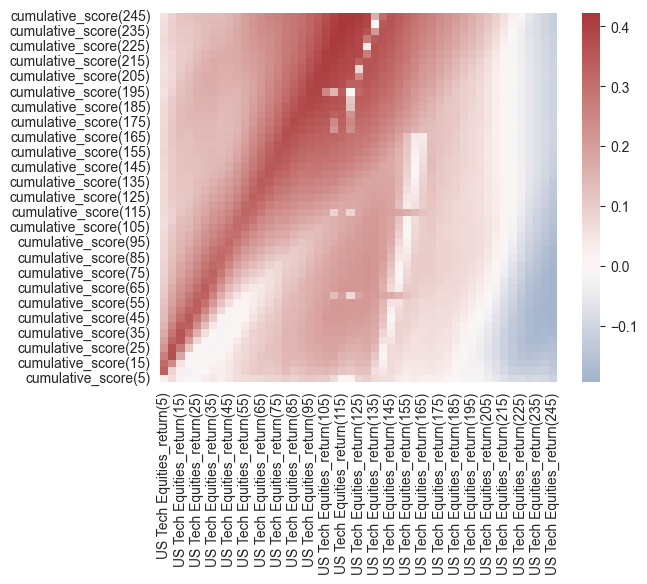} }}%
    \caption{Mitigated Spearman correlation between USTech and the cumulative sentiment score}\label{fig:USTech_Spearman_mitig}%
\end{figure}

\begin{figure}[!htbp]
    \centering
    {{\includegraphics[scale=0.65]{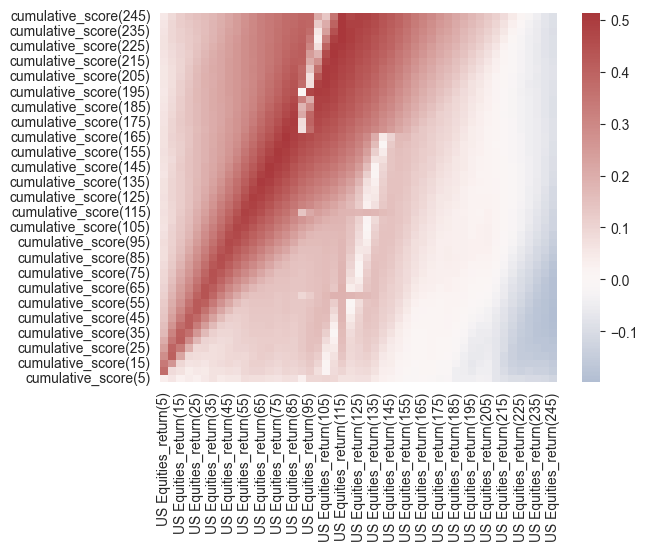} }}%
    \caption{Mitigated Spearman correlation between the U.S. and the cumulative sentiment score}\label{fig:US_Spearman_mitig}%
\end{figure}

\begin{figure}[!htbp]
    \centering
    {{\includegraphics[scale=0.65]{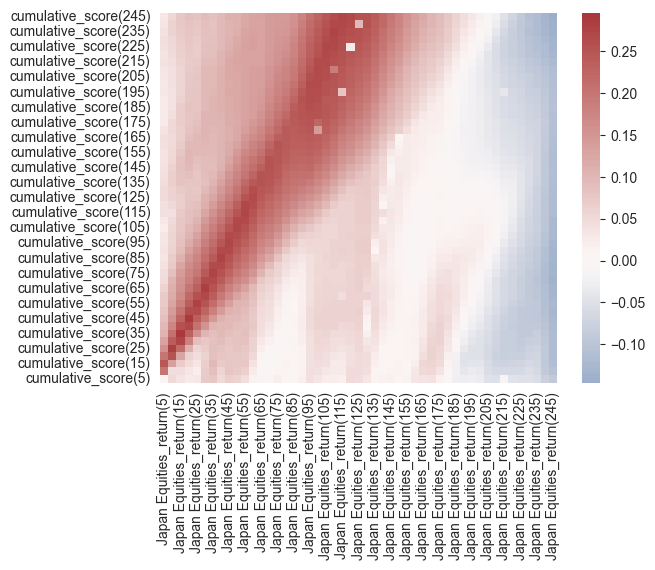} }}%
    \caption{Mitigated Spearman correlation between Japan and the cumulative sentiment score}\label{fig:Japan_Spearman_mitig}%
\end{figure}

\begin{figure}[!htbp]
    \centering
    {{\includegraphics[scale=0.65]{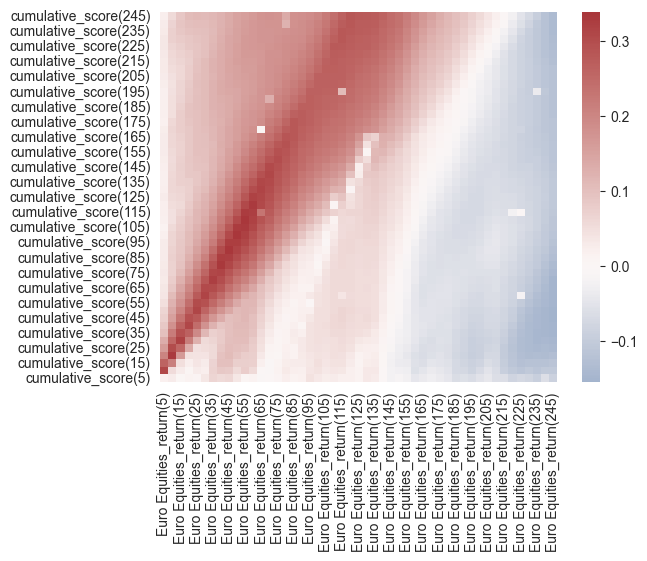} }}%
    \caption{Mitigated Spearman correlation between Euro and the cumulative sentiment score}\label{fig:Euro_Spearman_mitig}%
\end{figure}

\begin{figure}[!htbp]
    \centering
    {{\includegraphics[scale=0.65]{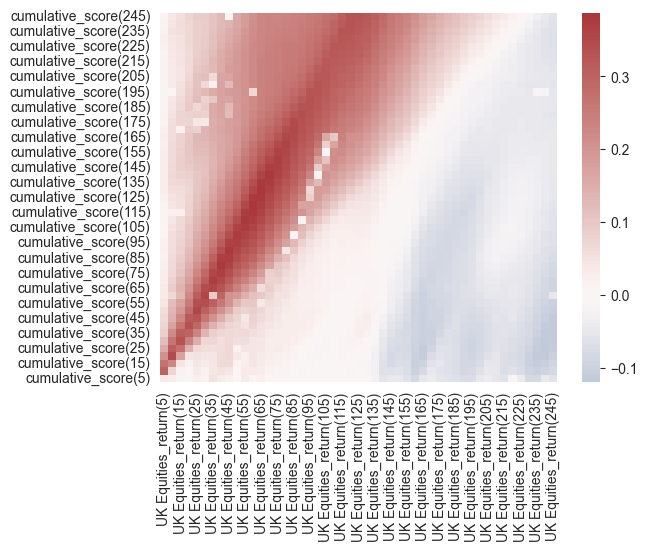} }}%
    \caption{Mitigated Spearman correlation between the UK and the cumulative sentiment score}\label{fig:UK_Spearman_mitig}%
\end{figure}

\begin{figure}[!htbp]
    \centering
    {{\includegraphics[scale=0.65]{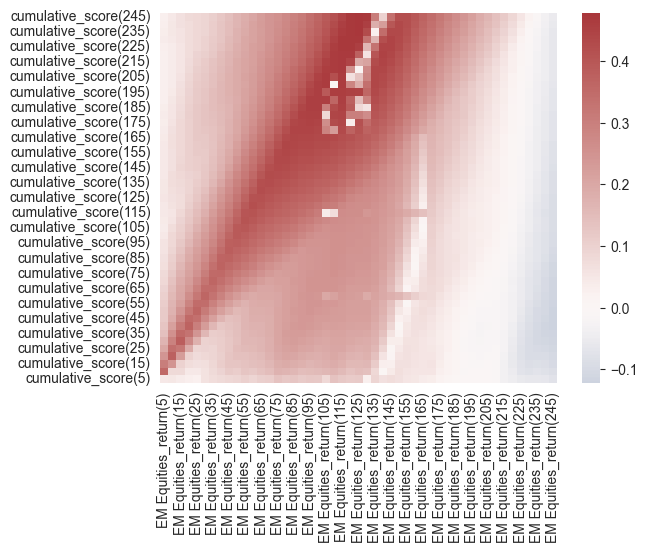} }}%
    \caption{Mitigated Spearman correlation between EM and the cumulative sentiment score}\label{fig:EM_Spearman_mitig}%
\end{figure}
\clearpage

\subsection{Optimal Point Determination for the Cumulative Score Lag-Value}
\begin{figure}[!htbp]
    \centering
    \includegraphics[scale=0.9]{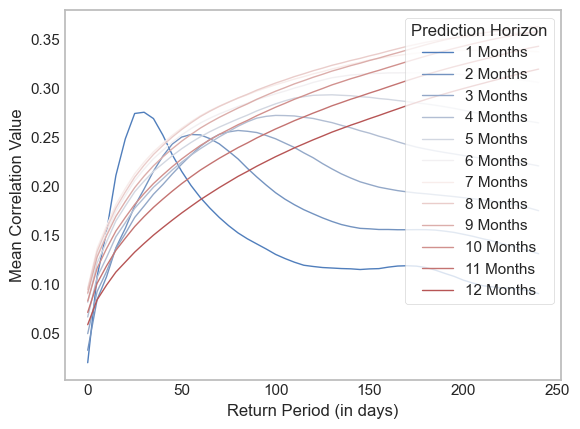}
    \caption{US Tech: Correlation of cumulative score lag over time.}\label{fig:tradeoff_USTech}%
\end{figure}

\begin{figure}[!htbp]
    \centering
    \includegraphics[scale=0.9]{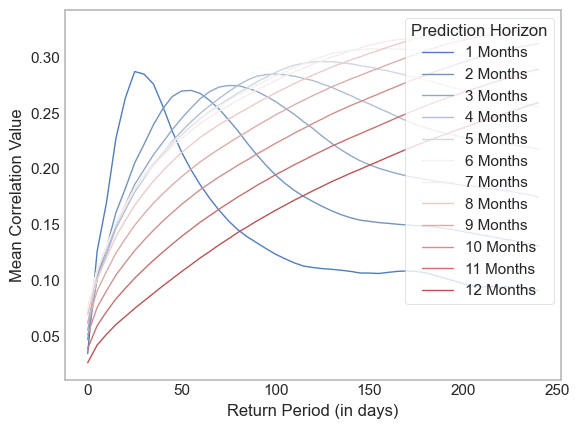}
    \caption{US: Correlation of cumulative score lag over time.}\label{fig:tradeoff_US}%
\end{figure}

\begin{figure}[!htbp]
    \centering
    \includegraphics[scale=0.9]{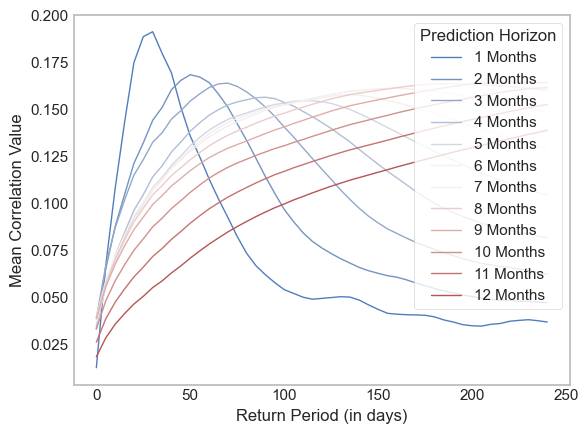}
    \caption{Japan: Correlation of cumulative score lag over time.}\label{fig:tradeoff_Japan}%
\end{figure}

\begin{figure}[!htbp]
    \centering
    \includegraphics[scale=0.9]{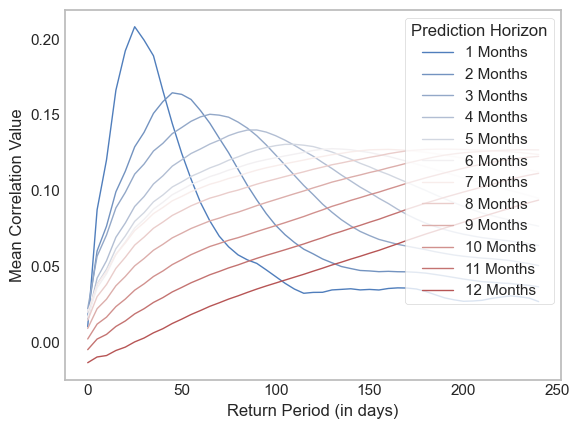}
    \caption{Euro: Correlation of cumulative score lag over time.}\label{fig:tradeoff_Euro}%
\end{figure}

\begin{figure}[!htbp]
    \centering
    \includegraphics[scale=0.9]{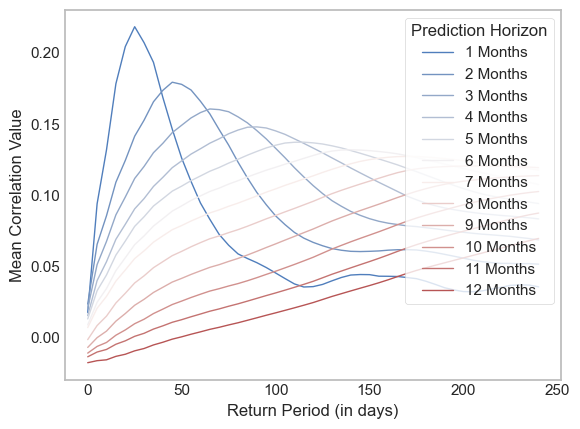}
    \caption{UK: Correlation of cumulative score lag over time.}\label{fig:tradeoff_UK}%
\end{figure}

\begin{figure}[!htbp]
    \centering
    \includegraphics[scale=0.9]{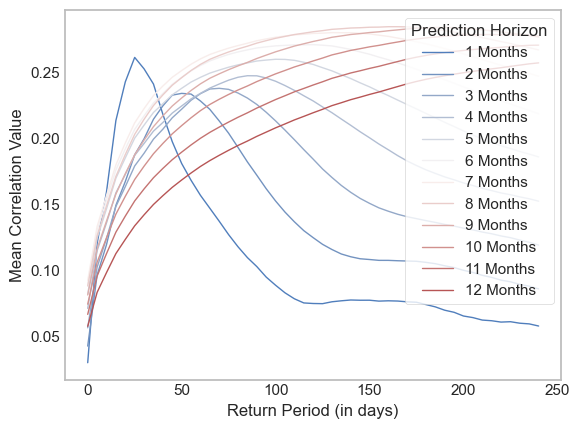}
    \caption{EM: Correlation of cumulative score lag over time.}\label{fig:tradeoff_EM}%
\end{figure}

\clearpage

\end{document}